\documentclass[prl,twocolumn,showpacs]{revtex4}
\usepackage{epsfig}
\begin{document}
%\pagestyle{empty}
%
%-- Define Page size
%
%\setlength{\topmargin}{0cm}
%\setlength{\headsep}{0in}
%\setlength{\textheight}{23cm}
%\setlength{\oddsidemargin}{0.4cm}
%\setlength{\marginparwidth}{.25in}
%\setlength{\textwidth}{16.0cm}
%\setlength{\parindent}{36pt}
%\baselineskip 12 pt
%
\title{
Photon Statistics for Single Molecule Non-Linear Spectroscopy 
}
\author{F. Shikerman, E. Barkai\\
Department of Physics,
Bar Ilan University, Ramat-Gan 52900 Israel
}
%\begin{document}
\begin{abstract}
{ 
 We develop the theory of non linear spectroscopy for a single
molecule undergoing stochastic dynamics and
interacting with a sequence of two laser pulses.
We find general expressions for the photon counting statistics  
and the exact solution to the problem for the Kubo-Anderson
process. 
In the limit of impulsive pulses the information on the photon statistics
is contained in the molecule's dipole correlation function. 
The selective  limit,
the semi-classical approximation, and the fast modulation limit,
exhibit rich general behaviors of this new type of spectroscopy. 
We show how the design of external fields leads to 
insights on ultra-fast
dynamics of individual molecule's which are different than those
found for an ensemble. 
}
\end{abstract}
%mean{SinMol ,Femto   ,Stoch.   ,Phot-Stat}
\maketitle

 Nonlinear optical interaction of a sequence of laser pulses
with matter provides a powerful tool for the investigation of
dynamics of ensembles of 
molecules in a wide variety of chemical, physical
and biological systems \cite{MukamelB}.
Recently,
van Dijk et al \cite{vanDijk}
reported of the first experimental study of an ultra-fast
pump-probe single molecule system.  Unlike the previous approaches
to non-linear spectroscopy where only the 
ensemble average response to the external fields is resolved, the
new approach yields direct information 
on single molecule dynamics, gained through the analysis of photon
counting statistics. 
Although the original experiment \cite{vanDijk}
was conducted on a molecule undergoing
a relatively simple relaxation process, the huge potential of combining
non-linear spectroscopy with single molecule spectroscopy,
inspires many unanswered theoretical questions:
What are the fundamental physical limitations of the investigation
of fast dynamics when spectral selectivity (defined below) is limited?
How does the information contained in these experiments
differ from the information contained in simpler 
continuous
wave experiments?  What is the finger print of 
coherence in these type of experiments, and precisely how
its influence on photon statistics is suppressed due to dephasing
processes? Finally, how to design the external control fields so that
new information on dynamics of molecules is gained. While the answer
to these questions, depends on the particular dynamics of the
molecule under investigation, 
we present a theory based on the Kubo-Anderson
model,
which
yields general insights on the problem.  

 Consider a sequence of two laser pulses interacting with a
single quantum system as a molecule, an atom, or a nano-crystal.
The pulses are assumed to be very short compared with the
radiative life-time of the emitter, so that the probability
of photon emission during a pulse is negligible, 
hence a pair of pulses yields two photons at most. 
 Repeating the experiment many times one may obtain the
probabilities $P_0,P_1$ and $P_2$ of emitting $0,1$ and $2$ photons.
In this manuscript we investigate the relation between $P_0,P_1$ and
$P_2$ 
and the dynamics of the underlying system  interacting with
the external fields. In turn this type of photon statistics
reveals important information on fast dynamics of molecules
in the condensed phase, information which is very difficult
to obtain using 
other theoretical
approaches to single molecule spectroscopy 
\cite{BarkaiPRL,Haw,Brown,Mukamel,BarkaiRev,Gopich}. 
We consider a model of a single molecule
undergoing spectral diffusion
and interacting
with  a pump-probe set up and 
show that  
depending on characteristics of the stochastic process  
and laser field parameters, 
different types of nonlinear spectroscopies emerge.
In particular, sensitivity to the phase accumulated by the molecule
in the delay period is found and impulsive and  selective
type of spectroscopies are considered in detail.
Spectral diffusion used here as a prototype of the molecule's  dynamics,
is found in many molecular systems \cite{BarkaiRev,YongPRL,Geva,Sanda},
is easy to detect using
the spectral trail technique, when the process is slow. 
Our goal is developing general methods suitable for the detection 
of a wider range of dynamics.

 The treatment begins with a two level molecule interacting
with a time dependent laser field according to the
optical Bloch equation \cite{MukamelB}
\begin{equation}
\dot{\sigma} = \hat{L}(t)\sigma + \hat{\Gamma} \sigma 
\label{eqBloch}
\end{equation}
where $\sigma = (\sigma_{\rm ee}, \sigma_{\rm gg}, \sigma_{\rm ge}, \sigma_{\rm eg})$.
As usual $\sigma_{\rm ee}$ and $\sigma_{\rm gg}$ are the populations of 
the excited and ground states, and 
$\sigma_{\rm ge},\sigma_{\rm eg}$ are the off
diagonal matrix elements of the density matrix.  
The operator
\begin{equation}
\hat{L}(t) =
\left(
\begin{array}{c c c c}
- \Gamma  & 0 & - i \Omega f(t)  & i \Omega f(t) \\
0 & 0 & i \Omega f(t) & - i \Omega f(t) \\
- i \Omega f(t) & i \Omega f(t) & i \omega(t) - \Gamma/2 & 0 \\
 i \Omega f(t) & - i \Omega f(t) & 0 & - i \omega(t) - \Gamma/2  
\end{array}
\right),
\label{eqOBE}
\end{equation}
describes the interaction of the molecule with the driving
electromagnetic field through 
$\Omega f(t)$ where $\Omega$ is the Rabi frequency, while 
$\omega (t)$ is the time dependent absorption frequency of the
molecule soon to be discussed. Finally the operator
%
%\begin{equation}
$\hat{\Gamma} =\Gamma |{\rm g}  \rangle \langle {\rm e}| $
%\left(
%\begin{array}{c c c c}
%0  & 0 & 0  & 0 \\
% \Gamma & 0 & 0  &  0 \\
% 0 & 0  & 0  & 0 \\
% 0 & 0  & 0  & 0 
%\end{array}
%\right).
%\label{eqLoOBE}
%\end{equation}
%
describes the transition  
from the excited state 
into the ground state,
due to 
spontaneous  emission with $\Gamma$ designating the emission rate.

 The spectral diffusion process 
$\omega(t)$ is modeled using the Kubo-Anderson
approach: 
$\omega(t)=\omega_0 + \delta w(t)$
where $\omega_0$ is the bare absorption frequency of the molecule,
and $\delta w(t)$ is a random function of time \cite{MukamelB,Kubo}.
We will assume that
this process is stationary, its mean is zero, its correlation
function is
$\langle \delta w(t_0+t) \delta w(t_0) \rangle=\nu^2 \psi(t)$ 
with $\psi(0)=1$  and $\psi(\infty)=0$. 
We will later demonstrate our results using the  Kubo-Anderson
process \cite{BarkaiRev,Berez},
where $\omega(t)=\omega_0 + \nu$ or $\omega(t)=\omega_0 - \nu$,
with the rate $R$ determining the transitions between the $+$
and $-$ states, which was used to model 
single molecules in low temperature
glasses \cite{Geva}.

 The photon statistics is obtained from the 
solution of Eq. 
(\ref{eqBloch}) in terms of an iterative expansion
in $\hat{\Gamma}$ \cite{Mukamel}.
This yields  $\sigma(t)=\sum_{n=0}^\infty \sigma^{(n)}(t)$ where
\begin{equation}
\sigma^{(n)}(t) = 
\int_{0} ^t{\rm d} \bar{t}_n  ... \int_{0} ^{\overline{t}_2} {\rm d}\bar{t}_1   {\cal G}(t, \bar{t}_n) \hat{\Gamma} ... {\cal G}(\overline{t}_2,\overline{t}_1) \hat{\Gamma} {\cal G}(\bar{t}_1,0)  \sigma(0).
\label{eqSig}
\end{equation}
The Green function
${\cal G} (t,t') = \hat{T} \exp\left[ \int_{t'} ^ t \hat{L} (\overline{t}) {\rm d} \overline{t} \right]$
($\hat{T}$ - the time ordering operator)
is the evolution operator
in the absence of the spontaneous emission
(i.e. without $\hat{\Gamma}$)
and therefore $\sigma^{(n)}(t)$ describes the state of the
system conditioned by $n$ emission events.  
The natural basis we shall
use is $| {\rm e}  \rangle=(1,0,0,0)$ which means that the molecule is in
the electronic excited state, $|{\rm g}   \rangle= (0,1,0,0)$ the ground state,
while $| {\rm c} \rangle=(0,0,1,0)$ and $| {\rm c}^{*}  \rangle= (0,0,0,1)$
describe the coherences.
Hence, the probability of emission of $n$ photons
is $P_n(t) =
 \langle {\rm e}| \sigma^{(n)} (t) \rangle + \langle  {\rm g} | \sigma^{(n)}(t) \rangle$,
which we calculate 
below in the limit of long measurement time
$t \to \infty$.

  In our model two identical pulses:  pump and  probe interact with the
molecule, 
\begin{equation}
f(t)= \left\{
\begin{array}{l l}
\cos \left( \omega_L t \right) & 0<t<t_1 \\
0 & t_1<t<t_2 \\
\cos \left[ \omega_L( t -t_2) \right] & t_2<t<t_3 \\
0  & t_3<t
\end{array}
\right.
\label{eqft}
\end{equation}
where $\omega_L$ is the laser frequency, $t_1=t_3-t_2=T$ are pulses duration,
and $\Delta=t_2 - t_1$ is the delay between the pulses. 
Since the Green function can be written as a product
of  propagators of shorter time intervals:
${\cal G} (t,0) =
{\cal G}(t,t_3) {\cal G} (t_3,t_2) {\cal G} (t_2,t_1) {\cal G} (t_1,0)$,
it is possible to treat everyone of them separately by dividing the
range of integration over time 
in Eq. (\ref{eqSig}),
and inserting closure relations
in appropriate places.
When the laser is turned off 
the Green functions ${\cal G}(t_2,t_1)$ and ${\cal G}(t,t_3)$
are nearly trivial, and one
may find general
expressions for $P_n$ (any $n$) for a particular realization
of the spectral diffusion process.  In the limit
of two short pulses obeying $\Gamma T\ll 1$  we neglect the processes
where photons are emitted during the pulse events, and find 
\begin{widetext}
\begin{equation}
P_n=P_n ^{{\rm Cla}} \left[\omega\left(t_1\right),\omega\left( t_2 \right)\right] +\left\{ e^{ i \theta \Delta - \Gamma \Delta/2 } A_n^{{\rm Coh}} \left[ \omega \left(t_1 \right), \omega \left( t_2 \right) \right] + \mbox{C.C.} \right\},
\label{eqPn1}
\end{equation}
here $\theta \Delta=\int_{t_1} ^{t_1 + \Delta} \omega(t) {\rm d} t$ is
the random phase accumulated in the delay interval. 
$P_n^{{\rm Cla}}$ given in Table 1 below, describe the semi-classical paths
in the sense that they include only the transitions between the
pure excited and ground states, whereas   $A_n ^{{\rm Coh}}$ represent
the contribution of the coherence to the photon statistics. 
\begin{tabular}{|c| c| c|}
\hline
$n$ & $P_n^{{\rm Cla}}$ & $A_n ^{{\rm Coh}}$ \\
\hline
$0$ & 
$ \langle {\rm g}| {\cal G}\left[\omega\left(t_2 \right) \right] | {\rm e} \rangle \langle {\rm e} | {\cal G}\left[ \omega(t_1) \right] | {\rm g}\rangle e^{ - \Gamma \Delta} +
\langle {\rm g}| {\cal G}\left[\omega(t_2)\right] | {\rm g} \rangle \langle {\rm g}|{\cal G}\left[\omega(t_1)\right]| {\rm g} \rangle  $ &
$ \langle {\rm g} | {\cal G}\left[\omega(t_2)\right] | {\rm c} \rangle \langle c | {\cal G}\left[\omega(t_1)\right] | {\rm g} \rangle $ \\
\ & \ & \\
\hline
$ 1$ &  
$\langle {\rm g} | {\cal G}\left[\omega(t_2)\right] | {\rm g} \rangle\langle {\rm e} | {\cal G}\left[\omega(t_1)\right]| {\rm g} \rangle \left( 1 - e^{ - \Gamma \Delta} \right) +
\langle {\rm e} | {\cal G}\left[\omega(t_2)\right]| {\rm e} \rangle \langle {\rm e} | {\cal G} \left[\omega(t_1)\right] | {\rm g} \rangle e^{ - \Gamma \Delta}
$
&
$ \langle {\rm e} | {\cal G}\left[\omega(t_2)\right] |c\rangle \langle c | {\cal G} \left[\omega(t_1)\right] | {\rm g} \rangle $ \\
\ &
$+ \langle {\rm e} | {\cal G}\left[\omega(t_2)\right]| {\rm g} \rangle \langle {\rm g} | {\cal G}\left[\omega(t_1)\right]| {\rm g} \rangle$
&
\ \\
\ & \ & \\
\hline
$ 2$ &  
$\langle {\rm e} | {\cal G}\left[\omega(t_2)\right] | {\rm g} \rangle \langle {\rm e} | {\cal G}\left[\omega(t_1)\right]| | {\rm g} \rangle\left(1 - e^{ - \Gamma \Delta} \right)$
&
$ 0 $ \\
& \ & \  \\
\hline
\end{tabular}
$$ \mbox{Table 1} $$
\end{widetext}
In our further calculations we assume that changes in the 
absorption frequency of the molecule during the pulse events are negligible,
namely that the rate of the spectral
diffusion process changes $R$ satisfies $R T \ll 1$. As a consequence
the random absorption frequency of the molecule at the moment of excitation
will be taken as $\omega(t)=\omega(t_1)$ during the first
pulse event and $\omega(t)=\omega(t_2)$ during the second pulse.
If this condition is met, the Green functions needed for the calculation
of the matrix elements in Table 1 are found within the rotating 
wave approximation (RWA) \cite{remark}
$
{\cal G}\left[\omega(t_1)\right]={\cal G}(t_1,0)=\exp\left\{ \hat{L}^{{\rm RWA}}[\omega(t_1)]T \right\}$ 
and similarly for ${\cal G}\left[\omega(t_2)\right]$,
\begin{equation}
\hat{L}^{{\rm RWA}}[\omega(t_1)] =
\left(
\begin{array}{c c c c}
-\Gamma  & 0 & {-i \Omega \over 2} & {i \Omega \over 2} \\
0 & 0 & {i \Omega \over 2} & {-i \Omega \over 2} \\
 {-i \Omega \over 2} & {i \Omega \over 2} & - {\Gamma \over 2} - i \delta(t_1) & 0 \\
 {i \Omega \over 2} & {-i \Omega \over 2} & 0 & - {\Gamma \over 2} + i \delta(t_1)
\end{array}
\right)
\label{eqRWA}
\end{equation}
where 
$\delta(t_1)=\omega_L-\omega(t_1)$ 
is the detuning.
Otherwise, the pulses yield time average information on the dynamics. 

%SHOULD WE INCLUDE THIS MATRIX:
%and the  RWA rotation matrix
%$A = | {\rm e} \rangle \langle {\rm e}| + |{\rm g}  \rangle \langle {\rm g} | + (e^{- i \omega_L T} |C\rangle \langle C | + e^{i \omega_L T} | {\rm c}^{*}\rangle \langle C^{*}|)$.
%

The results in Table 1 
describe the possible physical paths of photon emission.
For example consider the first term
of $P_0^{{\rm Cla}}$ in Table 1, 
the molecule starts in
the electronic ground state $| {\rm g} \rangle$, it then evolves with the
Green function of the first pulse ${\cal G}\left[\omega(t_1)\right]$ 
without emitting
a photon to the excited state $|{\rm e} \rangle$, it remains in the excited
state in the delay period without emitting a photon 
[with probability $\exp(-\Gamma \Delta)$],
and then the second pulse stimulates the molecule from
$|{\rm e} \rangle$ to $| {\rm g}\rangle$. 
From Eq. (\ref{eqPn1}) we see that the non-classical terms $A_n^{{\rm Coh}}$
are important only when $\Gamma \Delta$ is not too large, as
expected.
Also $A_2 ^{{\rm Coh}}=0$ in Table 1, 
since the emission of
two photons from two short pulses is possible only when one photon
is spontaneously 
emitted in the time interval between the pulses and the second photon after 
the second pulse, hence in this case the coherence 
is lost due to the collapse of the wave function. 

When $\Gamma T \to 0$ 
normalization condition
is $P_0+P_1+P_2=1$ as expected.
In this case the spontaneous emission does not contribute during
pulse events, and reversibility and
symmetry of matrix
elements is found: 
$\langle {\rm e} |{\cal G}[\omega(t_1)]|{\rm e} \rangle=
\langle {\rm g} | {\cal G}[\omega(t_1)]| g\rangle$,
$\langle {\rm e} |{\cal G}[\omega(t_1)]|{\rm g} \rangle=\langle {\rm g} |{\cal  G}[\omega(t_1)]| {\rm e}\rangle$ etc.
It is then easy to use Table 1 and show that 
$P_0 ^{{\rm Cla}}[\omega(t_1),\omega(t_2)]+
P_1 ^{{\rm Cla}}[\omega(t_1),\omega(t_2)] + P_2 ^{{\rm Cla}}[\omega(t_1),\omega(t_2)]=1$,
so the classical paths conserve probability, 
and $A_1 ^{{\rm Coh}}=-A_0 ^{{\rm Coh}}$. These simple relations
are valid for any realization of the underlying stochastic
process and they are valuable in reducing the number of matrix
elements needed for the calculation of the photon statistics
for a specific stochastic path.  

 To obtain the solution of the problem one must take averages
of $P_n$ in  Eq. 
(\ref{eqPn1}) over the stochastic process.
This requires knowledge of the joint probability density
function (PDF) $P[\omega(t_1),\omega(t_2),\theta \Delta]$ 
of finding the molecule's absorption frequency in the infinitesimal
range near $\omega(t_1)$ at $t_1$, near $\omega(t_2)$ at $t_2$
with accumulated random phase  
$\theta \Delta$.
Before considering the concrete example of the Kubo Anderson process,
we consider different limits allowing us to make a number of conclusions
valid for all types of spectral diffusion processes. 

{\em Impulsive limit $\nu\ll\Omega$. --}
Taking the limit $T \to 0$, $\Omega \to \infty$ 
in the pulse Green function Eq. 
(\ref{eqRWA})
while $\Omega T$ remains finite 
in such a way that $\nu \ll \Omega$ we define the impulsive limit. 
The matrix elements of ${\cal G}[\omega(t_1)]$ and ${\cal G} [ \omega(t_2)]$
{\em become independent of the value of $\omega(t)$ at the moment
of the excitation}, provided that the laser detuning is small compared with
$\Omega$.
Thus instead of the multi variable PDF  
$P[\omega(t_1),\omega(t_2),\theta \Delta]$ we now have to deal
only with the one variable PDF of the phase $\theta \Delta$. 
As a result the photon statistics shows an interesting relation
with linear continuous wave spectroscopy.  Using Eqs. (\ref{eqPn1},
\ref{eqRWA}) and Table 1  we find for stationary processes
\begin{widetext}
$$ \lim_{\Omega \to \infty,T \to 0}  \langle P_0 \rangle  =
 e^{-\Gamma \Delta} \sin^4\left( { \Omega T \over 2} \right)  + \cos^4 \left( { \Omega
T \over 2} \right) - {1 \over 2} e^{ - \Gamma \Delta/2} \sin^2 \left( \Omega T \right)
{\rm Re}\left[ \phi(\Delta) e^{ i \omega_0 \Delta} \right],
$$
\begin{equation}
\lim_{\Omega \to \infty,T \to 0} \langle  P_1 \rangle = {1 \over 2} \sin^2 \Omega T \left\{ 1  +  e^{ - {\Gamma \Delta \over 2} } {\rm Re} \left[\phi(\Delta) e^{ i \omega_0 \Delta} \right]\right\}, \ \ \ 
\lim_{\Omega \to \infty,T \to 0}  \langle P_2 \rangle= \left(1 - e^{ - \Gamma \Delta}\right) \sin^4 { \Omega T \over 2}
\label{Eqpooo}
\end{equation}
\end{widetext}
where 
\begin{equation}
\phi(\Delta) = \langle \exp( i \int_0 ^\Delta \delta \omega(t') {\rm d} t' \rangle
\label{eqCor}
\end{equation} 
is the well investigated Kubo-Anderson correlation function 
whose Fourier transform is the line shape of the 
molecule according to the Wiener--Khintchine theorem \cite{Kubo}. 
This type of impulsive 
limit offers no spectral resolution in the sense that
the photon statistics Eq. (\ref{Eqpooo})
is sensitive only to the random phase
$\theta$
accumulated in the time interval between the pulses 
and not to the temporal state of the molecule at the
time of the pulses. 
From Eq. (\ref{Eqpooo}) we see that 
for a $\pi/2$ pulse with $\Omega T=\pi/2$ 
the importance of the coherence terms 
and hence the correlation function $\phi(\Delta)$, 
on 
the photon statistics is the
strongest, since the $\pi/2$ pulse excites the off diagonal terms. 

{\em Semi-classical approximation.--} The influence of coherence on photon
statistics in many experimental cases is expected to be
 difficult to detect.
Either due to dephasing effects or simply because of the large value
of the optical transition frequency. In these cases a practical approximation
is to remove all the coherence terms, and leave only the semi-classical
terms $P_n ^{{\rm Cla}}$. From Table 1 it is easy to see that in this
case we must deal with the two dimensional PDF
$P[\omega(t_1),\omega(t_2)]$, 
instead of $P[\omega(t_1),\omega(t_2),\theta \Delta]$ 
for the general case. And then
\begin{equation}
\langle P_n \rangle = \langle P_n ^{{\rm Cla}} \rangle=
\int_0 ^\infty \int_0 ^\infty P_n ^{{\rm Cla}} P[\omega(t_1),\omega(t_2)]
{\rm d} \omega(t_1) {\rm d} \omega(t_2).
\end{equation}
Below we give an explicit example for this case for the
two state Kubo-Anderson model.

% We now obtain the averages $\langle P_0 \rangle,\langle P_1 \rangle$
%and $\langle P_2 \rangle$   
%with respect to the stochastic spectral diffusion
%process.
%The probabilities in Eq.
%(\ref{Eqpooo})
%depend only on a single random variable: the phase $\theta$, through
%trigonometric  $\exp(i \theta \Delta)$ terms, hence their averages
%will depend on expressions of the type $\langle \exp(i \theta \Delta)\rangle$.
%For stationary processes with the mentioned $\pi/2$ pulses
%we find 
%%
%\begin{equation}
%\langle P_0 \rangle= {1\over 4} + {e^{- \Gamma \Delta}\over 4}  -{ 1 \over 2} e^{ - \Gamma \Delta/2} {\rm Re}\left[ e^{i (\omega_0 \Delta + \omega_L T)} \phi\left( \Delta \right) \right]
%\label{eqp0}
%\end{equation} 
%%
%\begin{equation}
%\langle P_1 \rangle = { 1 \over 2} \left\{ 1 +  e^{ - \Gamma \Delta/2} {\rm Re} \left[ e^{i (\omega_0 \Delta + \omega_L T)} \phi\left( \Delta \right)  \right] \right\}
%\label{eqp1}
%\end{equation} 
%
%with the correlation function 
%\cite{remark}
%
%\begin{equation}
%\phi(\Delta) = \langle \exp( i \int_0 ^\Delta \delta \omega(t') {\rm d} t' \rangle.
%\label{eqCor}
%\end{equation} 
%%
%These general expressions, which are valid for any stationary process,
%show a surprising relation to  continuous wave spectroscopy. 
%The correlation function Eq. (\ref{eqCor}) is the well investigated 
%Kubo-Anderson correlation function 
%whose Fourier transform is the line shape of the 
%molecule according to the Wiener--Khintchine theorem \cite{Kubo}. 
%We now go beyond the impulsive limit. 

{\em Exact solution.--} 
Now we obtain the exact solution for the two state Kubo-Anderson Poissonian
process, where the absorption frequency of the molecule
jumps between a $+$ and $-$ states where $\omega=\omega_0\pm \nu$.
We denote the initial state, during the first pulse with $i=+$ or
$i=-$, similarly the final state of the molecule at the second pulse
is $f=+$ or $-$. Since the random phase $\Delta \theta = \omega_0 \Delta + \nu( T^{+} - T^{-})$ where $T^{\pm}$ are occupation times in
states $+$ and $-$ \cite{Berez}, satisfying $\Delta =T^{+} + T^{-}$,
the joint PDF $P(i,f,\Delta \theta)$ is equivalent to the joint PDF
$h(i,f,T^{+})$ of finding the molecule
in state $i$ in the first pulse, 
state $f$ in the second and with the occupation
time  $T^{+}$ between the two pulses.
We leave technical details on the calculation of $h(i,f,T^{+})$
for a longer publication,  
taking averages over Eq.
(\ref{eqPn1}) gives:
\begin{widetext} 
\begin{equation}
\langle P_n \rangle = \sum_{i=\pm,f=\pm} {\cal P}^{i f} P_n ^{{\rm Cla}} \left[\omega\left(i\right),\omega\left( f \right)\right] +e^{- \Gamma \Delta/2 }\left\{ e^{i \left( \omega_0 - \nu\right)\Delta}\hat{h}\left(i,f,- 2 i \nu \right)  A_n^{{\rm Coh}} \left[ \omega \left(i\right), \omega \left(f\right) \right] + \mbox{C.C.} \right\},
\label{eqPn11}
\end{equation}
where $\hat{h}\left(i,f,- 2 i \nu \right)$ is the Laplace $T^{+} \to 2 i \nu$
transform of $h(i,f,T^{+})$
\begin{equation}
\hat{h}(\mp,\pm,-2 i \nu) 
= e^{ - \Delta( R - i \nu) } { \sinh\left[ \Delta \sqrt{ R^2 - \nu^2} \right] R \over 2 \sqrt{ R^2 - \nu^2}}, \ \ \
%%\label{eqPn12F}
%%\end{equation}
%
%% \begin{equation}
\hat{h}(\pm,\pm, - 2 i \nu) 
=
{e^{- \Delta(R - i \nu)} \over 2} 
\left[ \cosh\left( \Delta \sqrt{ R^2 - \nu^2} \right) \pm { i \nu \sinh \left( \Delta \sqrt{ R^2 - \nu^2} \right) \over \sqrt{ R^2 - \nu^2} } \right]
\label{eqPn13F}
\end{equation}
\end{widetext}
while ${\cal P}^{\pm \pm}=[1+\exp(- 2 R \Delta)]/4$, 
${\cal P}^{\pm \mp}=[1-\exp(- 2 R \Delta)]/4$, are probabilities of
finding the particle initially in state $i$ and finally in state $f$,
which are easy to obtain from Poissonian statistics.

As an illustration we apply Eq. 
(\ref{eqPn11}) to the case of two infinitely strong
$\Omega T =\pi$ pulses, so no coherence is built. At first
assume that $\Omega\ll \nu$
and $\omega_L=\omega_0 + \nu$ so that the laser is in resonance
with the $+$ state and not with the $-$ state.
With this selective limit unlike the impulsive limit, temporal
resolution is found 
\begin{equation}
 \langle P_0 \rangle = {1 \over 4} \left( 1 + e^{ - 2 R \Delta} \right) \left( 1 + e^{ - \Gamma \Delta} \right),
 \ \ \ \langle P_1 \rangle = {1 \over 2} \left( 1 - e^{ - 2 R \Delta} \right) 
\label{eq23}
\end{equation}
and $\langle P_2 \rangle = 1 - \langle P_0 \rangle - \langle P_1 \rangle$.
These results make perfect physical sense, because the  molecule
emits a single photon only when it is once in state $+$ and once in state $-$,
hence
$\langle P_1 \rangle = {\cal P}^{+-} + {\cal P}^{-+}$. These simple solutions
show that for weak fields $\Omega \ll \nu$ the photon statistics
does not depend on frequency shifts $\nu$, and hence to explore
the dynamics of the molecule we must consider stronger fields.
Using Eq. (\ref{eqPn11}) for $\pi$ pulses, we find
the exact expression 
\begin{equation}
\langle P_1 ^{{\rm Cla}} \rangle = { 4 \nu^2 + \Omega^2 \cos^2 \left[ \sqrt{1 + 
\left( {2 \nu \over \Omega} \right)^2} { \pi \over 2} \right] \over 4 \nu^2 + \Omega^2 } \label{eq24} 
\end{equation}
$$ \left\{
{1-e^{-2R\Delta}\over 2} + {1 + e^{- 2 R \Delta} \over 2}
{ \Omega^2 \over \Omega^2 + 4 \nu^2} \sin^2 \left[\sqrt{1+({2 \nu \over\Omega})^2} {\pi \over 2} \right] \right\}  $$
which reduces to Eq. (\ref{eq23}) in the limit $\Omega \ll \nu$.
 In the opposite
limit $\Omega \gg \nu$ we have $\langle P_1 \rangle = 0$. 
Hence for the investigation of spectral shifts and rates 
we cannot use neither too weak or too strong
fields which with the condition $\Omega T =\pi$ means 
that pulses must not be chosen
arbitrarily short.
Eq. (\ref{eq24}) shows precisely what fields
yield information on the process beyond simple limits.

 {\em Fast modulation.--} An interesting case is the fast modulation
limit $R\gg \nu$, where motional narrowing effects take place.
We consider the limit $R \to \infty$ and
$\nu \to \infty$, in such a way the the spectral
diffusion dephasing rate $\Gamma_{{\rm SD}}=\nu^2/R$ 
is finite. The latter is a measurable physical
observable since it gives the width of the line shape in continuous wave
spectroscopy \cite{Kubo}.
To obtain the photon statistics in this limit we can 
proceed by expanding the exact
solution Eq. (\ref{eqPn11})
in terms of the large parameters $R$ and $\nu$. However, there is
a more general and simpler approach. Notice that  $\nu \ll R \ll 1/T$
and to have a finite probability of photon emission  $\Omega T$ 
must remain a constant of order unity, therefore
$\nu \ll \Omega$. Hence in this fast modulation limit the pulses
must be impulsive. Then Eq.
(\ref{Eqpooo})
holds with 
$\phi(\Delta) = \exp( - \Gamma_{{\rm SD}} \Delta/2)$, which
means that $(\Gamma_{{\rm SD}} + \Gamma)/2$ is the renormalized  
decay rate which damps the coherence terms.
This result is valid for many types of fast spectral diffusion
processes, and is not limited to the exactly solvable
two state process. 

%  In this limit we find
%using Eqs. (\ref{eqPn11F}-\ref{eqPn13}) 
%%
%\begin{widetext}
%\begin{equation}
%\langle P_n \rangle = {1 \over 4} \sum_{i=\pm,f=\pm}\left\{
%P^{{\rm Cla}}_n\left[\omega(i),\omega(f)\right] +\left( e^{i \omega_0 \Delta- \left(\Gamma_{{\rm SD}} + \Gamma\right) \Delta/2} A_n ^{{\rm Coh}}\left[\omega(i),\omega(f)\right] + \mbox{C.C.} \right) \right\}.
%\label{eqFast}
%\end{equation}
%%
%\end{widetext}
%%
%This equation shows that the role of the spectral diffusion rate
%$\Gamma_{{\rm SD}}$ is 
%to damp exponentially the effect of the coherence terms on the photon
%statistics. 
%For a $\pi/2$ pulse, with $\omega_L=\omega_0 + \nu$ 
%$$\langle P_0 \rangle = {9 + e^{-\Gamma \Delta} \over 16} -
%{e^{ - (\Gamma_{{\rm SD}} + \Gamma) \Delta/2} \over 8} \cos \left( \omega_0 \Delta + \omega_L T \right) $$
%\begin{equation}
%\langle P_1 \rangle = {1 \over 8}\left[ 3 + e^{ - \left( \Gamma_{{\rm SD}} + \Gamma\right) \Delta /2} \cos\left(\omega_0 \Delta +\omega_L T\right) \right].
%\end{equation}

{\em Semi-classical selective limit.--}  Neglecting the coherence terms
and taking a laser in resonance with the $+$ state and out
of resonance with the $-$ state i.e. $\omega_L = \omega_0 + \nu$,
$ \Omega\ll \nu$ we find from the exact result
Eq. (\ref{eqPn11})
$$ \langle P_0 \rangle = \cos^2 \left( { \Omega T \over 2} \right) +
\sin^4 \left( { \Omega T \over 2} \right) \left( 1 + e^{ - \Gamma \Delta} \right) {1 + e^{ - 2 R \Delta } \over 4}, $$
\begin{equation}
 \langle P_1 \rangle = \sin^2 \left( { \Omega T \over 2} \right) - \sin^4 \left({\Omega T \over 2} \right) {1 + e^{ - 2 R \Delta } \over 2} 
\label{eqClSel}
\end{equation}
and $\langle P_2 \rangle = 1 - \langle P_1 \rangle - \langle P_0 \rangle$.
Note that these results exhibit Rabi oscillations
and that $\langle P_1 \rangle$ is independent of
$\exp( - \Gamma \Delta)$ the latter behavior is general: 
for any spectral diffusing process $\langle P_1 ^{{\rm Cla}} \rangle$
is independent of $\exp( - \Gamma \Delta)$ in the limit of short and
strong pulses due to the mentioned symmetries of the matrix
elements of the pulse Green function. 

{\em Summarizing perspectives.--}
Theoretical investigation of the new field of single
molecule non-linear spectroscopy was presented. 
We showed that in the limit of impulsive pulses information
on the photon statistics is given by 
the Kubo-Anderson correlation function, for any stationary spectral diffusion
process. To obtain information on spectral shifts and spectral rates beyond
the trivial limits,  
one must choose carefully the duration of pulse and its strength,
as our analytical solutions clearly demonstrated. 
Our results can be checked in experiments in low temperature
glasses, 
and our methods  provide the theoretical basis for
the investigation of fast dynamics on the single molecule
level, thus opening the door for a vast number of applications. 

{\bf Acknowledgment} This work was supported by 
the Israel Science Foundation.


\begin{thebibliography}{99}

\bibitem{MukamelB} 
S. Mukamel {\em Principles of Nonlinear Optical Spectroscopy} Oxford 
Univ. Press. Oxford (1995). 
%\bibitem{Mollow} B. R. Mollow, {\em Phys. Rev. A}  {\bf 12}, 1919 - 1943 (1975)

% Single Molecule Pump Probe Detection Resolves Ultrafast Pathways
% in Individual and Coupled Quantum Systems. 
\bibitem{vanDijk} Erik M.H.P van Dijk et al {\em Phys. Rev. Lett.}
{\bf 94} 078302 (2005).

%{\em Time-Dependent Fluctuations in Single Molecule Spectroscopy: A Generalized
%Wiener--Khintchine Approach}
\bibitem{BarkaiPRL}  E. Barkai, Y.J. Jung, and R. Silbey,
{\em Phys. Rev. Lett.} {\bf 87}, 207403 (2001).

% Probing single-molecule dynamics photon by photon
\bibitem{Haw} H. Yang, X.S. Xie 
{\em  J. of Chem. Phys.} {\bf  117} 10965 (2002). 

\bibitem{Brown} Y. Zheng, F. L. H. Brown {\it Phys. Rev. Lett.}  {\bf 90}
238305  (2003).

%PHOTON STATISTICS: NONLINEAR SPECTROSCOPY 
\bibitem{Mukamel} S. Mukamel
{\em Phys. Rev. A} {\bf 68} 063821 (2003).

%{\em Theory of Single Molecule Spectroscopy: Beyond the ensemble Average}
\bibitem{BarkaiRev} E. Barkai, Y. Jung and  R. Silbey
{\em Annual Review of Physical Chemistry} {\bf 55}, 457 (2004).

%Theory of photon statistics in 
%single-molecule Forster resonance energy transfer
\bibitem{Gopich} 
I. Gopich, A. Szabo
{\em  J. of Chemical Physics} {\bf  122} 014707 (2005). 

% Title: Correlations in single molecule photon statistics: Renewal indicator
%\bibitem{Cao} J.S. Cao
%{\em  J. of  Phys. Chem. B} {\bf  110}  19040 (2006)


%Theory of single-molecule optical line-shape distributions in low-temperature glasses
\bibitem{Geva} E. Geva, J.L. Skinner
{\em  J. of Phys. Chem. B} {\bf 101} 8920 (1997). 

%Influence of Spectral Diffusion on
\bibitem{YongPRL} 
Y. He, E. Barkai
{\em Phys. Rev. Lett.} {\bf 93} 068302 (2004).

%Liouville-space pathways for spectral diffusion in photon statistics from single molecules 
\bibitem{Sanda} F. Sanda, S. Mukamel
{\em Phys. Rev A} {\bf  71} 033807 (2005). 

\bibitem{Kubo} R. Kubo, M. Toda, and N. Hashitsume {\em Statistical Physics 2}
Springer Berlin (1995).

%\bibitem{Toyo} R. Kubo, and Y. Toyozawa, {\em Prog. Theoret. Phys.} {\bf 13} 160
%(1955).  

%Theory of single-molecule fluorescence spectroscopy of two-state systems
\bibitem{Berez}  A.M. Berezhkovskii, A. Szabo, G.H Weiss 
{\em J. of Chem. Phys} {\bf  110} 9145 (1999)

\bibitem{remark} In a future publication we will discuss the influence
of the laser phase on the photon statistics, briefly this yields
phase shifts in the coherence terms.  

%\bibitem{Katz} For example dynamics of Josephson junctions qubits, 
%N. Katz et al {\em Science}  {\em 312} 1498 (2006) since the influence
%of coherence
%was already detected in microwave experiments, or three level molecules,
%etc. 


%
%\bibitem{Old Text}
%LD TEXT. From the general solution Eq. (\ref{eqPn1}) we see that the probabilities
%P_n$ depend on three random variables: the absorption
%requency at the first pulse, $\omega(t_1)$, at the second pulse
%\omega(t_2)$, and the phase $\theta$. From Table 1 it is
%lear that $P_n ^{{\rm Cla}}$ depend only on the random temporal state
%f the molecule at the time of the pulses $\omega(t_1)$ and $\omega(t_2)$
%ut not on the phase $\theta$, while the amplitude $A_n ^{{\rm Coh}}$ depends
%n all the three random variables.
%he joint probability density function
%p[\omega(t_1),\omega(t_2),\theta]$  
%f finding the system in state $\omega(t_1)$
%nitially, $\omega(t_2)$ in final pulse, and that the phase is $\theta$,
%s therefore needed for the calculation of averages  
%f photon number probabilities. 
%We now obtain an exact solution for the problem
%or  the two state Kubo-Anderson
%rocess where the molecule can be either in the plus $\omega(t)=\omega_0+\nu$
%or minus $\omega(t)=\omega_0 - \nu$ states.
%e denote the initial state the molecule, during the first pulse with
%i=+$ or $i=-$ and the final state with $f=+$ or $-$.
%We notice that the phase of the two
%tate Kubo-Anderson model is given by $\theta \Delta= \omega_0 \Delta + 
%nu(T^{+} - T^{-})$ where $T^{\pm}$ is the total occupation time
%n the states $+$ and $-$ during the delay interval $\Delta=T^{+}+T^{-}$.  
%revious work of Berezhkovskii et al \cite{Berez} showed how
%o obtain distribution of the occupation time $T^{+}$.
%n the quantum problem under consideration 
%he distribution of $T^{+}$ which starts in state $+$ or $-$ and
%nds in state $+$ or $-$ are calculated, since the starting point
%nd the ending point of the molecule as well as the phase accumulated
%n the delay interval are needed for the calculation of
%he photon statistics.  We find
%
%begin{widetext} 
%begin{equation}
%langle P_n \rangle = \sum_{i=\pm,f=\pm} {\cal P}^{i f} P_n ^{{\rm Cla}} \left[\omega\left(i\right),\omega\left( f \right)\right] +e^{- \Gamma \Delta/2 }\left\{ \hat{p}^{if}\left(\Delta \right)  A_n^{{\rm Coh}} \left[ \omega \left(i\right), \omega \left(f\right) \right] + \mbox{C.C.} \right\},
%label{eqPn11a}
%end{equation}
%
%begin{equation}
%hat{p}^{\pm \pm}(\Delta) = {1 \over 2} \left[ \cosh\left( \Delta \sqrt{ R^2 - \nu^2} \right) \pm {i \nu \over \sqrt{R^2 - \nu^2}} \sinh \left( \Delta\sqrt{R^2 - \nu^2} \right) \right] e^{ i \omega_0 \Delta- \Delta R}
%label{eqPn12}
%end{equation}
%begin{equation}
%hat{p}^{\pm \mp}(\Delta) = {1 \over 2} {R \over \sqrt{R^2 - \nu^2}} \sinh\left( \Delta \sqrt{ R^2 - \nu^2} \right)  e^{ i \omega_0 \Delta - R \Delta}
%label{eqPn13}
%end{equation}
%end{widetext}
%hile ${\cal P}^{\pm \pm}=[1+\exp(- 2 R \Delta)]/4$, 
%{\cal P}^{\pm \mp}=[1-\exp(- 2 R \Delta)]/4$, are probabilities of
%inding the particle initially in state $i$ and finally in state $f$,
%hich are easy to obtain from Poissonian statistics. 

%Stochastic Lionville, Langevin, Fokker-Planck, and master equation approaches 
%to quantum dissipative systems
%Author(s): Tanimura Y (Tanimura, Yoshitaka)
%Source: JOURNAL OF THE PHYSICAL SOCIETY OF JAPAN 75 (8): 
%Art. No. 082001 AUG 2006 

\end{thebibliography}
\end{document}